\documentstyle[12pt,aaspp4, tighten]{article}

\begin{document}

\title{Looking for Distributed Star Formation in L1630: \\
       A Near-infrared ($J$, $H$, $K$) Survey}

\author{Wenbin Li\altaffilmark{1,4} and Neal J. Evans II\altaffilmark{2,4}}
\affil{Department of Astronomy, The University of Texas \\
       Austin, TX 78712}
\authoremail{wenbin, nje@astro.as.utexas.edu}

\author{Elizabeth A. Lada\altaffilmark{3,4}}
\affil{Astronomy Department, The University of Florida\\
       Gainesville, FL 32608}
\authoremail{lada@astro.ufl.edu}

\altaffiltext{1}{Electronic mail: wenbin@astro.as.utexas.edu}
\altaffiltext{2}{Electronic mail: nje@astro.as.utexas.edu}
\altaffiltext{3}{Electronic mail: lada@astro.ufl.edu}
\altaffiltext{4}{Visiting Astronomer, Kitt Peak National Observatory, National
     Optical Astronomy Observatories, which is operated by the
     Association of Universities for Research in Astronomy, Inc.\
     (AURA) under cooperative agreement with the National Science
     Foundation.}

\vskip5cm
\centerline{\it To Appear in}
\centerline{\it The Astrophysical Journal}
\centerline{\it October 1997 Vol. 488}

\clearpage

\begin{abstract}

We have carried out a simultaneous,
multi-band ($J$, $H$, $K$) survey over an area of 1320 arcmin$^2$
in the L1630 region, concentrating on the region away from the dense molecular
cores and with modest visual extinctions ($\leq 10$ mag).  Previous studies
found that star formation in L1630 occurs mainly in four localized clusters,
which in turn are associated with the four most massive molecular cores
(Lada et al.\ 1991; Lada 1992).
The goal of this study
is to look for a distributed population of pre-main-sequence
stars in the outlying areas outside the known star-forming cores.
More than 60\% of the pre-main-sequence stars
in the active star forming regions of NGC 2024 and NGC 2023
show a near-infrared excess in the color-color diagram.
In the outlying areas of L1630, excluding the known
star forming regions, we found that among 510 infrared sources with the
near-infrared
colors ($(J-H)$ and $(H-K)$) determined and photometric uncertainty
at $K$ better than 0.10 mag,
the fraction of the sources with a near-infrared excess is 3\%--8\%;
the surface density of the sources with a near-infrared
excess is less than half of that found in the distributed population in L1641,
and 1/20 of that in the young cluster NGC 2023.
This extremely low fraction and low surface density of sources
with a near-infrared excess
strongly indicates that recent star formation activity
has been very low in the outlying region of L1630.  The sources without
a near-infrared excess could be either background/foreground field stars,
or associated with the cloud, but formed a long time ago
(more than 2 Myrs).
Our results are consistent with McKee's model of photoionization-regulated
star formation.

\end{abstract}


\section{Introduction}

Observations have established that star formation in our Galaxy
occurs mainly in molecular clouds.
However, we do not understand what conditions are required for
molecular clouds to form stars.
A naive argument based on Jean's criterion would predict that most of the
molecular clouds in the Galaxy are unstable,
resulting in a star formation rate of $> 130$--$400$
$M_{\sun}~{\rm yr}^{-1}$ (e.g., Evans 1991). But this number is about two
orders of magnitude higher than the average Galactic star-formation
rate over the last few Gyrs (e.g., Scalo 1986).
It is generally believed that magnetic fields play a crucial role
in supporting the clouds.

McKee (1989) proposed a model of photoionization-regulated star formation
to explain the cloud support and low star-formation
rate (also see Bertoldi \& McKee 1996 for a more elaborate model based on
similar ideas).  This model combines the idea of magnetic support with the
observation that photon-dominated regions (PDR's) occupy a large fraction
of molecular gas.  In McKee's scheme, magnetically supported clumps rely on
ambipolar diffusion to get rid of magnetic flux (thus magnetic support).
In the outer layers of molecular clouds, where the column density
is relatively low, the interstellar far-ultraviolet (FUV) radiation is able
to maintain a relatively high ionization fraction. As a result,
the timescale for ambipolar diffusion is too long to allow efficient
star formation.
Thus, McKee's theory would predict little or no recent star formation
in the outer layers of molecular clouds.

At a distance of about 400 pc (Anthony-Twarog 1982; Brown,
de Geus, \& de Zeeuw 1994), the Orion molecular cloud complex is
one of the best laboratories for a study of star formation.
The L1630 cloud is in the northern part of the Orion complex, including
the well-known nebulae NGC 2024 and NGC 2023.  In this cloud,
most young stars are found in four clusters.
Specifically, Lada et al.\ (1991) argued that 58\%--96\% of the young stellar
objects in L1630 are formed in these clusters, depending on the correction
for the field stars.
Furthermore, the four clusters are associated with the four
largest and most massive cores (Lada 1992).
This result appears to be consistent with McKee's model of
photoionization-regulated star formation. According to McKee's model,
star formation happens only in the cores with
visual extinction (equivalent to column density) above a threshold value
$A_V^{th}$:
above $A_V^{th}$, cosmic rays are the dominant source of
ionization and the ionization fraction is low;
below this threshold,
the interstellar FUV radiation ionizes the gas to a higher degree, resulting
in slower ambipolar diffusion.
Therefore the outer layers of the clouds with $A_V < A_V^{th}$ have unfavorable
conditions for stars to form, because the
ambipolar diffusion is too slow for
cores to collapse in a reasonable time ($\le 10^8$ yr).
The exact value of $A_V^{th}$ ranges between 4--8 mag,
depending on the local FUV intensity.

Interestingly, in L1641---part of the Orion complex and several degrees south
of L1630---Strom, Strom, \& Merrill (1993a) found
a population of isolated pre-main-sequence (PMS) stars
distributed throughout the cloud (thus called the ``distributed population'').
Based on follow-up spectroscopic studies,
Allen (1996) claimed that roughly 50\% of the stars in L1641 (exclusive
of OMC1) are formed in
isolation, called the ``distributed formation'' mode, in contrast to the
``cluster formation'' mode found in L1630 by Lada et al.\ (1991).

The L1630 and L1641 clouds provide an excellent opportunity for a comparison
study, since they are close together and similar
in many aspects, including mass and size.
We want to know if stars are truly formed differently in these two clouds.
The original survey of Lada et al.\ (1991) still leaves room for
a possible distributed population of young stars,
because of the unknown contribution by field stars.
If L1630 and L1641 do have different modes of star formation, we want to
know how the physical and chemical conditions, and the
external conditions in these two clouds differ,
and whether these differences cause their different star formation patterns.
The knowledge will help us understand star formation and the
support and evolution of molecular clouds.

The experiments in L1630 by Lada et al.\ (1991) and that in L1641 by
Strom et al.\ (1993a) differ considerably.
The survey of L1630 was carried out in the $K$ band only, with a
5 $\sigma$ detection limit of 14 mag, while that
of L1641 was conducted in three bands ($J$, $H$, $K$), to the limits of
16.8, 15.8, and 14.7 mag respectively (5 $\sigma$).
The two surveys covered about the same area (0.7 square degrees),
but the L1630 study covered areas both with and without strong CS emission,
while the L1641 study
did not include the extremely rich and active OMC1 molecular core.
These differences make a direct comparison difficult.

Aiming to resolve these issues, we carried out a new near-infrared (NIR)
survey in L1630, using the large-format
detectors of the SQIID camera to obtain simultaneous, multi-band
($J$, $H$, $K$) observations toward L1630.  We designed the
experiment to concentrate on regions with modest extinction
($A_V \leq 10$ mag)---comparable to that of the survey regions of Strom et al.\
(1993a) in L1641 (also Allen 1996).  According to McKee's model
of photoionization-regulated star formation, clouds with $A_V \le A_V^{th}$
have unfavorable conditions for forming stars, and thus evidence of
recent star formation would test the fundamental idea of the model.
The goals were to search for star formation in
the regions away from the dense cores in a systematic manner and
to look carefully for evidence of star formation in the distributed mode.
Recent, distributed star formation can be found
by examining the NIR colors, since many PMS stars, such as
T Tauri stars and Herbig Ae/Be stars, show NIR excess emission.
However, this method will not identify older PMS stars, since they may have
lost their circumstellar disks, or whatever causes strong NIR excess.
Evidence that older PMS stars formed in the distributed mode can be found
by looking for an enhancement of source density.
This study needs a good knowledge of the distribution of field stars.
We selected large, off-cloud, control fields in different directions for this
purpose.

In this contribution, we report the results of the NIR survey and the study
of the NIR colors.  Observations and data reduction
are summarized in \S 2.  In \S 3, results of the survey are presented,
followed by the study of the NIR colors for sources in the control
fields and the on-cloud fields in \S 4.
We discuss implications of the results in \S 5.
Finally in \S 6, a summary is given.  The study of star counts will
be reported in a follow-up paper, where we will compare the observed
stellar surface density in the cloud against predictions
obtained from the field stellar density and an extinction estimate.

\section{Near-Infrared Observations \& Data Reduction}

\subsection{NIR Survey Area}

The survey area is marked in Figure~\ref{survey-area} as the box overlaid on
an image of the region retrieved from the on-line service of the
Digitized Sky Survey\footnote{URL: {\tt
http://www-gsss.stsci.edu/dss/dss\_form.html}}.
The total area on the cloud is about 1320 arcmin$^2$.

\placefigure{survey-area}

The study covered several known objects, including the reflection
nebula NGC 2023, part of another reflection nebula IC 435
(roughly 20\arcmin ~east of NGC 2023), and the
Horsehead nebula.  The survey extended westward beyond the
photoionization front that marks the edge of L1630.

An area of 5\farcm8 $\times$ 5\farcm8 centered on NGC 2024
(5h39m12s, $-01$\arcdeg 55\arcmin 42\arcsec) (Lada et al.\ 1991)
was also repeatedly imaged
throughout the run with 5 sec integrations,
for the purpose of checking the pointing.

In addition, three off-cloud, but nearby, control fields were selected
to be free of molecular gas based on the $^{12}$CO $J=1\rightarrow 0$ map
in Maddalena et al.\ (1986).
The total area of 533 arcmin$^2$ of the control fields is
considerably larger than that of the original survey of L1630
(376 arcmin$^2$) by Lada
et al.\ (1991) or of L1641 (87.5 arcmin$^2$) by Strom et al.\ (1993a).
To assess the contribution from background and foreground stars,
it is very important to have large control fields in different directions,
especially considering the possible complications caused by the Orion OB1
association in the region.
The information about the control fields is summarized
in Table~\ref{control-fields}.  The (0, 0) reference position is
that of NGC 2024 (Lada et al.\ 1991).  For comparison, the crude coordinates
of NGC 2023, which is more or less in the middle of the whole survey area,
are also listed (ref. Figure~\ref{survey-area}).

\subsection{Observations and Data Reduction}

The NIR data were obtained on 1995 February 3--6,
using the Simultaneous Quad Infrared Imaging Device (SQIID, Ellis et al.\
1992; Merrill 1995a) on the Kitt Peak National Observatory 1.3-meter
telescope.  The camera used four $256\times 256$ PtSi arrays to
record simultaneous images in the $J$, $H$, $K$, and $L$ bands.
For this project, only data in $J$, $H$, and $K$ were taken.
The pixel sizes were 1\farcs395 ($J$), 1\farcs366 ($H$), and
1\farcs361 ($K$),
giving roughly 5\farcm8 $\times$ 5\farcm8 fields of view (Merrill 1995a).

All four nights were photometric, with seeing of 3\arcsec\ or better
throughout the run.
Ten or 11 NIR standard stars from Elias et al.\ (1982)
were observed throughout each night.
The standards were selected to cover the widest
possible ranges of $(J-H)$ and $(H-K)$ colors. They were imaged at
a range of airmasses comparable to those for L1630.
At least one standard star was observed every two hours.
For all the fields, dithered pairs of images ($15\arcsec$ E--W shifts) with
180 sec integrations were taken.  For most fields, frames of
shorter integrations were also taken to obtain photometry of
bright sources, which could saturate the detectors in 180 sec.
The dark current was measured at the beginning and the end of each night.
A globular cluster (M3) was imaged with 60 sec integration to
extract coordinate transformations between different bands.

Primarily, IRAF and the SQIID Package (Merrill 1995b) were used for data
reduction and photometry.
We applied the standard procedures of CCD image reduction, including
non-linearity correction, dark current subtraction, sky subtraction,
and flat fielding.  The DAOFIND task in IRAF was used to extract stars,
with a detection threshold of 3.7--4.0 $\sigma$ (standard deviation
of the background).  All images were also visually examined to include
bright sources missed by the DAOFIND routine.  We used the APPHOT task
to obtain photometry.
The final photometry was based on the averaged images of dithered pairs
with a total integration time of 360 sec.
Photometric transformations were obtained from the standard
stars, including atmospheric correction and color terms.  The results
were applied to the object stars to calibrate them to the CIT system.

\section{Results}

The typical photometric uncertainty is tabulated in Table~\ref{phot-error}.
The 90\% completeness limit of the survey is 16.5 mag at $J$, 15.5 mag at $H$,
and 14.5 mag at $K$ respectively.
The 90\% completeness limit was set
where the number of sources detected ($\ge 3.7~\sigma$) in only one frame of
the dithered pair
but not the other one reaches 10\% of the number of sources detected
in both frames.

Eight frames of short integrations on NGC 2024 were combined to produce an
image with a total integration of 40 sec on the cluster.
The effective area of the
frame is 21.28 arcmin$^2$, and the 90\% completeness limits are about
14.4 mag ($J$), 13.6 mag ($H$), and 12 mag ($K$).

In the survey area of 1320 arcmin$^2$ on the L1630 cloud,
the source numbers detected at $J$, $H$, and $K$
to the 90\% completeness limits are 1353, 1244, and 919 respectively,
while in three off-cloud
control fields with a total area of 533 arcmin$^2$, there are 919,
655, and 404 sources detected down to the 90\% detection limits in the
three bands respectively.

Twenty-five HST guide stars were identified from the NIR images and
were used to calibrate the astrometry.
The absolute astrometric errors of equatorial positions of
sources are less than $4\arcsec$ for right ascension and smaller than
$1\arcsec$ for declination.

The sources detected in the survey are displayed in Figures~\ref{starmapj},
{}~\ref{starmaph}, ~\ref{starmapk}.  Note that two off-cloud fields are also
plotted below the on-cloud fields for the convenience of visual
comparison; they are actually much farther away.

\placefigure{starmapj}

\placefigure{starmaph}

\placefigure{starmapk}

\section{Near-infrared Colors}

The NIR color-color diagram ($(J-H)$ vs. $(H-K)$) is a useful
tool for identifying young stellar objects and probing their circumstellar
material (e.g., Lada \& Adams 1992).  Circumstellar matter, such as an
accretion disk, produces thermal emission at long wavelengths,
resulting in excess infrared emission.
Since newly-born stars may
have circumstellar accretion disks and/or circumstellar
envelopes (Beckwith \& Sargent 1993; Hartmann, Kenyon, \& Calvet 1993),
an NIR color excess is a good indicator of young age.

In the rest of \S 4, we will look for PMS stars in the outlying
region of L1630 by studying the NIR colors.
First we will examine the colors of the background/foreground
stars in the off-cloud control fields (\S 4.1).  Then we will study
the colors of PMS stars (\S 4.2).
With the characteristics of the NIR colors of field stars and PMS stars
established,
we will examine the region of L1630 away from the dense cores (\S 4.3).
In \S 5, we will compare L1630 and L1641 and discuss implications of the
results.

\subsection{Field Stars in the Control Fields}

Figure~\ref{cc-empty1} displays the loci of main-sequence (MS) and
giant stars in the
NIR color-color diagram. The reddening band, using the reddening law of
Rieke \& Lebofsky (1985, RL85), is also marked to illustrate the location that
MS and giant stars move to when suffering extinction.

\placefigure{cc-empty1}

There is no obvious difference in the color-color diagram between
different control fields. So we combined the three fields
and plotted the sources in Figure~\ref{cc-off},
overlaid upon the loci of MS and giant stars and the reddening band.
The left panel includes stars with $K$ photometric
uncertainty no larger than 0.10 mag, and the right panel includes stars with
$K$ uncertainty up to 0.20 mag.

\placefigure{cc-off}

The left panel of Figure~\ref{cc-off} clearly shows that the sources
in the control fields indeed follow the loci of MS
and giant stars closely, except that about a half dozen sources ($\sim 2.3\%$)
are shifted along the reddening band, indicating some
apparent reddening.  These sources could be attributed to poor photometry,
or extinction of $A_V < 5$ mag.
However, they will not have any significant
impact on the statistical results of the whole sample.
In the right panel of Figure~\ref{cc-off}, we see a larger scattering of
sources around the loci of
MS and giant stars compared with the left panel, and many more sources
fall to the right of the reddening band.  Within the range of photometric
uncertainty, most of these sources are still consistent with being
MS and giant stars suffering minimal extinction,
but the large dispersion could complicate the analysis.
Therefore we choose to use only stars with $\Delta K \leq 0.10$ mag for all the
analysis of color-color diagrams.

Figure~\ref{cc-off-cont} uses contours to illustrate
the distribution of stars in the control fields quantitatively.
The agreement between the ``ridge'' of the contours and the loci of MS
and giant stars is excellent.
The field stars are peaked at two
locations: ($(H-K)$, $(J-H)$) = (0.09, 0.35), and (0.20, 0.64).
They correspond to the spectral types of K0V and M2V/K4III respectively.
In addition, there is no evidence for contamination by the Orion OB1
association, since we do not see any source concentration
along the reddening line from the O, B, A type stars.
We conclude that
the control fields are very good and indeed do not suffer obvious extinction
and that most sources detected in the control fields are field MS and
giant stars.

\placefigure{cc-off-cont}

\subsection{PMS Stars in the Active Star Forming Regions}

It is known that star formation is active around the reflection nebula
NGC 2023, where a cluster of PMS stars has been identified
(DePoy et al.\ 1990; Lada et al.\ 1991).
Therefore we separate the NGC 2023 region from the rest of the survey area.
A rectangle
($-160\arcsec \leq \Delta \alpha \leq 140\arcsec$,
$-1490\arcsec \leq \Delta \alpha \leq -1240\arcsec$)
is defined by visual examination to cover the NGC 2023 cluster,
as marked out in Figures~\ref{starmapj}--\ref{starmapk}, with the
same total area as Lada et al.\ (1991) measured using quantitative criteria.

In Figure~\ref{cc-clusters} we present the color-color diagrams of the
young clusters associated with NGC 2024 and NGC 2023.
Note that the data of NGC 2024 only cover about 4\farcm6~$\times$~4\farcm6
centered at the (0, 0) position (specified for Figure~\ref{survey-area})
and have a lower sensitivity than the data of other regions in our survey.
The actual size of the NGC 2024 cluster is about 180 arcmin$^2$
(Lada et al.\ 1991).

\placefigure{cc-clusters}

{}From Figure~\ref{cc-clusters}, the ranges of NIR colors for the
sources in NGC 2024 and 2023
are wide and comparable, roughly 0.8--2.6 for $(J-H)$
and 0.2--1.8 for $(H-K)$.
Taking $(H-K)$, its range corresponds to $A_V$ up to 25 mag,
if attributed to interstellar extinction and using the extinction law of RL85,
but excess emission also affects the colors.

Many sources in the two young clusters clearly fall to the right of the
reddening band of field stars, indicating an NIR excess.
For NGC 2024, 15 out of 42 sources
fall to the right of the reddening band; for NGC 2023,
eight out of 11 are to the right of the reddening band.

To assess quantitatively how many sources falling to the right of the
reddening band can be attributed to photometric errors, we took the
following approach:
we compute the distance of each
source (with $\Delta K \leq 0.10$ mag) to the right-most reddening line
(that of O8V stars) and compare the distance with its
uncertainty (1 $\sigma$) due to photometric errors.
For the control fields, among the 16 sources to the right of the reddening
band, only one star has a distance more than 1 $\sigma$,
and no star has a distance more than 2 $\sigma$;
for the clusters, 6 sources (out of 8) in NGC 2023 and 13 sources (out of
15) in NGC 2024
are more than 1 $\sigma$ away from the reddening band.  Therefore,
in the clusters, the sources to the right of the
reddening band are mainly due to NIR excess emission, while
in the control fields, all sources to the right of the reddening band
can be attributed to photometric errors.  Since some stars with
NIR excess will be shifted to the left by errors, and since most stars
have intrinsic colors closer to the left side of the reddening band
(ref. Fig.~\ref{cc-off-cont}), this approach is
a very conservative way of calculating the fraction of NIR-excess
sources, giving only a lower limit. In this section, we use the total number
of sources to the right of the reddening band, and
admit that a small fraction of sources could be displaced to the right
of the reddening band simply by photometric errors (in the control fields,
it is 7\%).

In NGC 2023, 73\% of the sources show an NIR excess, and in NGC 2024
the fraction is 36\%.
Our result on NGC 2023 is higher than the 50\%
found by DePoy et al.\ (1990).
This discrepancy could be attributed to different detection limits and the
slightly different area coverage.
For NGC 2024, Comer\'on et al.\ (1996) and Lada et al.\ (1997) both found
the fraction of sources showing an NIR excess
to be greater than 60\%.  The lower fraction
we obtained in NGC 2024 is probably caused by our worse sensitivity.
Therefore, we conclude that the fraction of NIR-excess sources in
the young clusters NGC 2024 and 2023 of L1630 is at least 60\%.
In the rich, young Trapezium cluster of L1641, McCaughrean and
collaborators (1995) found that 60--80\% of the stars have infrared excess
emission.  This number agrees well with our result on NGC 2023.

\subsection{The Outlying Region in L1630}

Figure~\ref{cc-ex2023} is the color-color diagram of sources in the outlying
region---the survey area excluding the NGC 2023 part, as defined in
Figure~\ref{starmapj}.
Evidently, only a few sources lie to the right of the reddening band.
Quantitatively, 42 out of the total of 510 sources (8\%) fall to the right
of the reddening band.  Furthermore, only 18 sources (3\%) are
more than 1 $\sigma$ from the reddening band.  Photometric errors
could in principle have shifted most of the 42 sources to the right
of the reddening band.  Based on the fraction of sources displaced to the
right of the reddening band by photometric errors in the control fields,
we would expect roughly 34 such sources in the outlying region in L1630.
Figure~\ref{cc-ex2023-cont} shows contours of the sources in the outlying
region.  Evidently these sources are well confined within the reddening
band.
\placefigure{cc-ex2023}
\placefigure{cc-ex2023-cont}
Therefore, the fraction of sources showing
NIR excess in the outlying region is 3\% to 8\% . This is
much lower than what we inferred for the active star forming regions NGC
2024 and NGC 2023.

The choice of reddening laws has little effect on our results.
Most other studies found $(J-H)/(H-K)$ less than that of RL85,
which would result in even fewer NIR-excess sources than we have
obtained using RL85.  The conversion between $E_{J-H}$ or $E_{H-K}$ to
$A_V$ would be slightly affected, with the tick marks in the color-color
diagrams moving up the reddening lines by less than 0.1 mag
(for $A_V = 5$ mag) for most other studies.

The extremely small fraction of NIR-excess sources in the outlying region
of L1630 indicates that the star formation activity there
is totally different from that in NGC 2024 or NGC 2023,
and that the sources detected in the outlying region
are qualitatively different from those in NGC 2024 and NGC 2023.
In particular, many of these sources may be background field stars. If these
are more easily seen in the outlying region, they could dilute
the population of embedded sources, partially explaining the low fraction
of NIR-excess sources. To avoid this problem, we also calculated the
surface density of NIR-excess sources.
In NGC 2023, the surface density of the sources with more than 1 $\sigma$
NIR excess is 0.286 stars/arcmin$^2$, and in the outlying areas of L1630,
the density is less than 0.014 stars/arcmin$^2$.  Thus, on this measure,
there is a factor of 20 difference between the cluster and the outlying
region.

\section{Discussion}

\subsection{PMS Stars vs.\ NIR Color Excess}

The NIR color excess cannot identify the complete population
of PMS stars (Aspin \& Barsony 1994; Aspin, Sandell, \& Russell 1994).
In young clusters like NGC 2024 and NGC 2023, the fraction of PMS
stars is more than 60\%; but based
on a sample of T Tauri stars and Class I sources (Lada and Wilking 1984;
Lada 1987), Aspin et al.\ (1994) found that roughly 60\% of them do {\it not}
show NIR excess and could be confused with reddened background MS or
giant stars.

There are three reasons why a PMS star may not display any NIR excess
on the color-color diagram.  The first is that some young
stars, e.g., many weak-line T Tauri stars (WTTS),
seem to be born without a source of excess emission.  They have been
called naked T Tauri stars (NTTS) (Walter 1986).

The second reason is detectability in the color-color diagram.
Even though a PMS star has a disk, its NIR excess may not be large
enough to shift the star to the right of the reddening band.
Lada and Adams (1992)
found that only half of the Class II sources (Lada and Wilking 1984;
Lada 1987) in Taurus (generally believed to have disks)
have large enough excess to be detectable in the color-color diagram.

The third reason why a PMS star may not show any NIR excess in the color-color
diagram is the evolutionary stage.  Circumstellar
accretion disks/envelopes are considered to be the primary sources of the
NIR excess.  When a PMS star evolves toward the MS, its disk/envelope
dissipates at some point, and consequently the star loses its NIR excess.
The lifetime of circumstellar disks is quite uncertain, ranging
from 2 to 10 Myrs (Lada \& Lada 1995; Strom, Edwards, \& Strutskie 1993).
Generally speaking, NIR excess indicates a young age.

\subsection{Is there Star Formation in the Outlying Region of L1630?}

There exist two plausible scenarios that could explain
the tiny fraction of NIR-excess sources in the outlying regions of L1630.
First, there is no significant embedded population in the outlying region and
the sources detected in the NIR are essentially all field MS and giant stars.
The second possibility is that there is a sizable population of
sources associated with the cloud in this region, but
these stars have formed a long time ago,
so that they do not have circumstellar disks substantial enough
to produce detectable NIR excess in the color-color diagram.
Their age must be older than the typical lifetime of NIR excess
attributed to circumstellar disks,
2--10 Myrs (Lada \& Lada 1995; Strom et al.\ 1993b).
As for the formation mode of this old population, we have no way of
constraining it, since a star at a distance of 400 pc would drift
43\arcmin ~in 5 Myrs, assuming a random velocity of 1 km/s.
This distance is much larger than the physical scale of a typical cluster,
and is comparable to our survey area.

Theoretically there exists a third possibility to explain the extremely
low fraction of NIR-excess sources in the outlying region of L1630.
If stars in the outlying region preferentially form without disks,
the low fraction of NIR-excess sources might be explained.
There is no obvious reason why this should be true, and we consider
it unlikely.

With the color-color diagram only, we cannot definitely determine the nature
of the sources detected in the outlying region of L1630 to
discriminate which of the first two scenarios is better.
In the follow-up paper, we will try to answer
this question by examining the stellar density in the whole survey area
in comparison to the surface density of the field stars in the off-cloud
control fields.  We will estimate the contribution of foreground stars
through the data and the Galactic models; then we will calculate the
expected stellar density through the cloud with the extinction estimate
from various molecular tracers.  If we find an enhancement of
surface density,
it would suggest an embedded population of ``older'' young stars that have
lost their NIR excess emission.

\subsection {Comparison to L1641}

Not only is the level
of star formation activity in the outlying regions of L1630 much lower
than that in the young clusters like NGC 2023, it
is also less than half of that in the outlying areas of L1641:
in L1641, the NIR survey of Strom et al.\ (1993a) found that in
the 2548 arcmin$^2$ outlying areas excluding
the clusters and aggregates,
33\% of 254 sources display a NIR excess of more than 1 $\sigma$.
Thus the surface density of the NIR-excess sources there is
0.033 stars/arcmin$^2$, more than twice the surface density in
the L1630 outlying regions.  The surface density in the aggregates in
L1641 is about 0.31 stars/arcmin$^2$, similar to what we find in NGC 2023,
so the difference between these two adjacent regions is
in the distributed population.

We will discuss the relative contribution to star
formation by young clusters and the distributed population more
quantitatively in the follow-up paper.

\subsection{McKee's Model}

Our results agree with the prediction of McKee's model of
photoionization-regulated star formation.  In the areas away from the dense
cores, low column density results in a high ionization fraction and a long
ambipolar diffusion time scale.  In these regions, the collapse time
is too long for stars to form efficiently.

Furthermore, as shown in Figures~\ref{cc-clusters}, \ref{cc-ex2023},
the sources in the outlying region have a
much smaller range of NIR colors than the PMS stars
in the active star formation regions NGC 2024 and NGC 2023.
For NGC 2024 and NGC 2023, more than 80\% of the sources have
$(H-K) >$ 0.6, while in the outlying region, 80\% of the sources have
$(H-K) \leq 0.6$.  Using the reddening law of RL85,
$(H-K) = 0.6$ corresponds to $A_V = 7$ mag.  If the reddening is not
local to each star, then the extinction of $A_V = 7$ mag is consistent
with McKee's prediction.

\section{Summary}

In this contribution, we have reported the results of a simultaneous,
multi-band ($J$, $H$, $K$) survey over an area of 1320 arcmin$^2$
in the L1630 region, concentrating on the outlying region with
modest visual extinctions ($\leq 10$ mag), as a complement to
the work on the dense cores and young clusters by Lada et al.\ (1991).
The goal was to look for a distributed population of young stars and
distributed star formation, as reported in the adjacent L1641 region
(Strom et al.\ 1993a; Allen 1996), and to test McKee's model of
photoionization-regulated star formation (1989).
In the on-cloud region, a total of 1353, 1244, and 919 NIR
sources were detected
to the 90\% completeness limits at $J$, $H$, and $K$ respectively.
In addition, three off-cloud control fields along different directions
were observed, with a total
area of 533 arcmin$^2$, and 919, 655, 404 sources were detected
in the three bands respectively.

Detailed study of the NIR colors for the off-cloud control
fields, the star-forming regions, and the outlying region
of L1630 indicates the following:
\begin{enumerate}
\item In the active star-forming regions of NGC 2024 and NGC 2023,
more than 60\% of the sources show an NIR excess in the color-color
diagram (($J-H$) vs. ($H-K$)).
\item Outside the known cluster NGC 2023, only a few sources show NIR excess
and most fall into the reddening band of field stars.
The  fraction of sources having NIR excess
is 3\%--8\%, much lower than what is found in active star-forming regions.
The surface density of sources with NIR excess in the outlying regions of
L1630 is less than half of that in L1641, and 1/20 of that in the young cluster
NGC 2023.
This low fraction and low surface density of NIR-excess sources
confirms the difference between L1630 and L1641.
\item The extremely small fraction of NIR-excess sources strongly suggests
that there is no recent, active star formation in the outlying region of L1630,
and that most sources detected
are either field stars, and/or stars formed from the cloud a long time ago.
\item The results in L1630 are consistent with the original work of Lada et al.
(1991) and the prediction of McKee's model of photoionization-regulated star
formation.
\end{enumerate}

\acknowledgments

We would like to extend our sincere thanks to the staff of KPNO,
especially Dr. Mike Merrill, for their excellent support.
We are also grateful to
Mr. Don Horner for helping with the observations.
We especially thank Dr. Dan Jaffe for stimulating discussions,
constructive suggestions, and a lot more.
Dr. Lori Allen deserves special credit for the challenging questions
and stimulating discussions in her referee's report.
This research has made use of
NASA's Astrophysics Data System Abstract Service,
the Simbad database, operated at CDS, Strasbourg, France,
and the On-line Digitized Sky Survey image retrieval service.
The Digitized Sky Survey was produced at the Space Telescope Science
Institute under U.S. Government grant
NAG W-2166. The images of these surveys are based on photographic
data obtained using the Oschin Schmidt
Telescope on Palomar Mountain and the UK Schmidt Telescope.
The plates were processed into the present
compressed digital form with the permission of these institutions.
The National Geographic Society - Palomar Observatory Sky Atlas (POSS-I)
was made by the California
Institute of Technology with grants from the National Geographic Society.
The Second Palomar Observatory Sky Survey (POSS-II) was made by the
California Institute of Technology
with funds from the National Science Foundation, the National Geographic
Society, the Sloan Foundation, the
Samuel Oschin Foundation, and the Eastman Kodak Corporation.
The Oschin Schmidt Telescope is operated by the California Institute of
Technology and Palomar Observatory.
The UK Schmidt Telescope was operated by the Royal Observatory Edinburgh,
with funding from the UK
Science and Engineering Research Council (later the UK Particle Physics
and Astronomy Research Council),
until 1988 June, and thereafter by the Anglo-Australian Observatory.
The blue plates of the southern Sky Atlas
and its Equatorial Extension (together known as the SERC-J),
as well as the Equatorial Red (ER), and the Second
Epoch [red] Survey (SES) were all taken with the UK Schmidt.
WL has been partially
supported by NSF grant AST-9317567,
a Continuing Fellowship and a David Bruton, Jr., Fellowship
of the University of Texas,
a Frank N. Edmonds,
Jr. Memorial Fellowship and a David Alan Benfield Memorial Scholarship
of the Department of Astronomy, the University of Texas.

\clearpage

\figcaption[figure/overlay-9.epsi]
{The NIR survey coverage is
overlaid on the image of the region taken from the Digitized Sky Survey.
The reference (0, 0) position is (5h39m12s, $-01$\arcdeg 55\arcmin 42\arcsec).
\label{survey-area}}

\figcaption[figure/starmapj.epsi]
{Source distribution
in the $J$ band.  Note that two off-cloud control fields are plotted for
comparison in the dotted-line boxes.  The central solid-line box
($-160\arcsec \leq \Delta \alpha \leq 140\arcsec$,
$-1490\arcsec \leq \Delta \alpha \leq -1240\arcsec$) defines the
young cluster NGC 2023 in a somewhat arbitrary way, but matches the total area
of the young cluster as Lada et al.\ (1991) measured with quantitative
criteria.
$J$ magnitudes are used to linearly size the dots in the plot
between $J = 10.5$ mag and $J = 16.5$ mag.
\label{starmapj}}

\figcaption[figure/starmaph.epsi]
{Source distribution in the $H$ band.
Similar to Figure~\ref{starmapj}, except that the dots are linearly scaled
by the $H$ magnitudes of sources between $H = 10.5$ mag and $H = 15.5$ mag.
\label{starmaph}}

\figcaption[figure/starmapk.epsi]
{Source distribution in the $K$ band.
Similar to Figure~\ref{starmapj}, except that the dots are linearly scaled
by the $K$ magnitudes of sources between $K = 10.5$ mag and $K = 14.5$ mag.
\label{starmapk}}

\figcaption[figure/cc-empty1.epsi]
{The NIR color-color diagram
of MS and giant stars (Koornneef 1983), and the reddening band
(Rieke \& Lebofsky 1985, RL85).  The crosses along the reddening lines are
marks of $A_V = 5, 10, 15, 20, 30$.
Since the calibration magnitudes of the NIR standards in Elias et al.\ (1982)
are in the CIT system, but the data in
Koornneef (1983) are closer to the Johnson system,
the following transformations have been used to
convert the Koornneef data to the CIT system:
$(J-H)_{CIT} = 0.92(J-H)_J - 0.005, ~(H-K)_{CIT} = 0.912(H-K)_J - 0.009$
(Bessell \& Brett 1988; Leggett 1992).
The same transformations have also been applied to the reddening law
of RL85 ($E(J-H)/E(H-K) = 1.698$), which is in the Johnson system.
In this work, all the NIR color-color diagrams
use the same MS/giant loci and reddening bands as in this figure.
\label{cc-empty1}}

\figcaption[figure/cc-off.epsi]
{The NIR color-color diagrams of the sources in the off-cloud control fields.
A total of 241 sources are plotted in the left panel, of
which the $K$ band photometric uncertainties are no bigger
than 0.10 mag, while in the right panel, 398 sources with the
$K$ band uncertainty up to 0.20 mag are plotted.
\label{cc-off}}

\figcaption[figure/cc-off-cont0.1.epsi]
{Contours of sources in the control fields in the color-color diagram.
It was obtained by sampling the distribution of sources in the left panel
of Figure~\ref{cc-off} with $0.05\times 0.05$ ($(J-H)\times (H-K)$)
boxes and half resolution steps.  The contours start from $1/(0.05^2)$
(number/mag$^2$), and increase by $2/(0.05^2)$ (number/mag$^2$) every level.
\label{cc-off-cont}}

\figcaption[figure/cc-clusters.epsi]
{The color-color diagrams of known star-forming regions NGC 2024 and
NGC 2023.  The data of NGC 2024 are from a coadded frame with
the total integration of 40 seconds, and the estimated 90\%
completeness limits are 14.4 mag ($J$), 13.6 mag ($H$), and 12 mag ($K$).
Forty-two sources with color measurements and $\Delta K \leq 0.10$ mag
are included for NGC 2024, and 11 sources for NGC 2023.
\label{cc-clusters}}

\figcaption[figure/cc-ex2023.epsi]
{The color-color diagram of sources in the outlying regions of L1630,
excluding NGC 2023, which is known to be star formation active.
A total of 510 sources are included.
\label{cc-ex2023}}

\figcaption[figure/cc-ex2023-cont0.1.epsi]
{Contours of the distribution of sources in Figure~\ref{cc-ex2023},
similar to Figure~\ref{cc-off-cont}.
\label{cc-ex2023-cont}}




\begin{deluxetable}{cccccc}
\tablewidth{0pc}
\tablecaption{The Off-cloud Control Fields for L1630. \label{control-fields}}
\tablehead{
\colhead{Field} & \colhead{Center $\Delta \alpha$\tablenotemark{a}}
                & \colhead{Center $\Delta \delta$\tablenotemark{b}}
                & \colhead{Area} & \colhead{$b$\tablenotemark{c}}
                & \colhead{$l$\tablenotemark{d}} \\
\colhead{} & \colhead{(\arcmin)} & \colhead{(\arcmin)}
           & \colhead{(arcmin$^2$)}
           & \colhead{(\arcdeg)} & \colhead{(\arcdeg)}}
\startdata
Control A & $-$57.2 & $-$13.3 & 12.8$\times$13.0 & $-$17.3 & 206.3\\
Control B & 91.0 & $-$67.3 & 13.1$\times$11.1 & $-$15.5 & 208.3 \\
Control C & 12 & 14.6 & 14.5$\times$15.3 & $-$15.0 & 204.4 \\
On-cloud (NGC 2023) & 0 & $-$21.7 & \nodata & $-$16.2 & 206.2 \\
\tablenotetext{a,b}{The reference position is $\alpha = $ 5h39m12s,
$\delta = $ $-$01\arcdeg 55\arcmin 42\arcsec ~(1950).}
\tablenotetext{c,d}{The galactic coordinates.}
\enddata
\end{deluxetable}


\begin{deluxetable}{cccc}
\tablewidth{0pc}
\tablecaption{Typical Photometric Errors of the NIR Survey. \label{phot-error}}
\tablehead{
\colhead{Mag} & \colhead{$J$} & \colhead{$H$} & \colhead{$K$} }
\startdata
11.5 & 0.02 & 0.02 & 0.02 \\
12.0 & 0.02 & 0.03 & 0.02 \\
12.5 & 0.02 & 0.03 & 0.04 \\
13.0 & 0.02 & 0.06 & 0.06 \\
13.5 & 0.03 & 0.06 & 0.10 \\
14.0 & 0.04 & 0.09 & 0.15 \\
14.5 & 0.05 & 0.11 & 0.23 \\
15.0 & 0.06 & 0.17 & 0.35 \\
15.5 & 0.10 & 0.23 & 0.52 \\
16.0 & 0.14 & 0.34 & 0.80 \\
16.5 & 0.21 & 0.51 & \nodata \\
17.0 & 0.32 & 0.84 & \nodata \\
17.5 & 0.49 & \nodata & \nodata \\
18.0 & 0.83 & \nodata & \nodata \\
\enddata
\end{deluxetable}
\clearpage

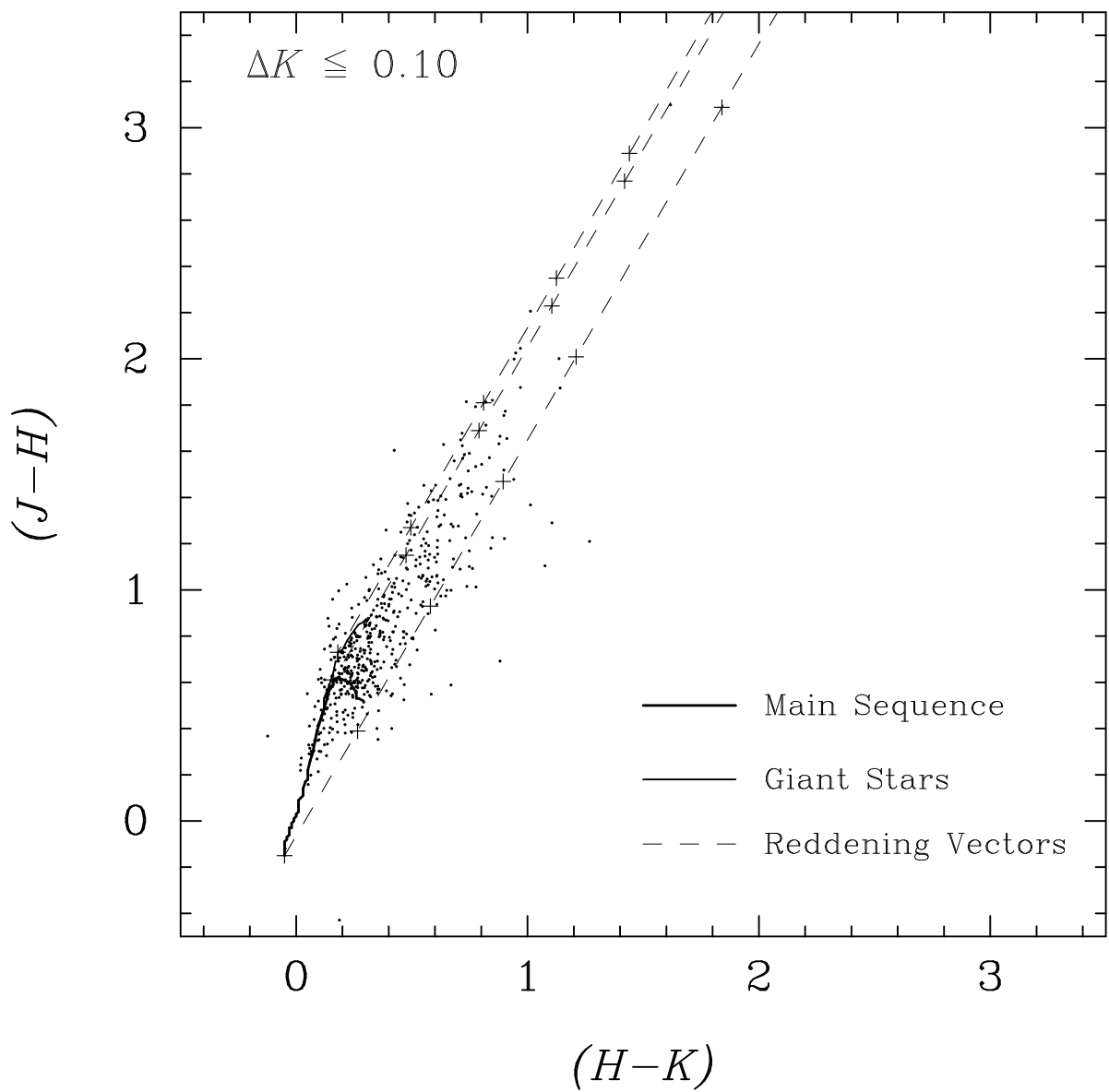
\end{document}